\documentstyle[12pt,epsfig]{article}
\begin{document}
\pagestyle{plain}

\def\ksi0{$\Xi_c^{0}$}
\def\ksip{$\Xi_c^{+}$}
\def\lam{$\Lambda_c^{+}$}
\def\om{$\Omega_c^{0}$}
\def\ee{ \end{equation} }
\def\be{ \begin{equation} }
\def\eea{ \end{eqnarray} }
\def\bea{ \begin{eqnarray} }

\newcommand{\figps}[5]{\begin{figure}
         \centerline{
         \epsfig{file=#1.ps,height=#2cm,width=#3cm,silent=}}
         \caption{#4} \label{#5}
                      \end{figure}}
\newcommand{\figeps}[5]{\begin{figure}
         \centerline{
         \epsfig{file=#1.eps,height=#2cm,width=#3cm,silent=}}
         \caption{#4} \label{#5}
                      \end{figure}}
\newcommand{\figgeps}[8]{\begin{figure}
         \centerline{
         \epsfig{file=#1.eps,height=#2cm,width=#3cm,silent=}
         \epsfig{file=#4.eps,height=#5cm,width=#6cm,silent=}}
         \caption{#7} \label{#8}
                      \end{figure}}

%
\setlength{\voffset}{-0.5in} \setlength{\textheight}{9in}
\setlength{\textwidth}{6.5in} \setlength{\oddsidemargin}{.01in}
\thispagestyle{empty}
\begin{flushleft}
\begin{small}
IRB-TH 1/97 \\
April 1997 \\
\end{small}
\end{flushleft}
\vspace{1.5cm}
\begin{center}
\begin{Large}
\begin{bf}
Inclusive Charmed-Baryon Decays and Lifetimes\\
\end{bf}
\end{Large}
\vspace{1.2cm}
\begin{Large}
Branko Guberina\footnote{E-mail: guberina@thphys.irb.hr} and 
Bla\v zenka Meli\'c
\footnote{E-mail: melic@thphys.irb.hr}\\
{\small Theoretical Physics Division, Rudjer Bo\v skovi\'c
Institute, P.O.B.1016\\
 HR-10001 Zagreb, Croatia\\}
\end{Large}
\end{center}
\vspace{3.5cm}
\begin{center}
{\bf Abstract}
\end{center}
\begin{quotation}
\noindent
We have quantitatively reanalyzed the inclusive charmed-baryon decays.
 New ingredients are the
Voloshin preasymptotic effects in semileptonic decays and the
Cabibbo-subleading
contributions to both semileptonic and nonleptonic decays. It has been found
that the Cabbibo-subleading Voloshin contribution essentially improves the
theoretical semileptonic branching ratio of $\Lambda_c^{+}$, in agreement with
experiment. The semileptonic branching ratios for $\Xi_c^{+}$ and
$\Omega_c^{0}$
are found to be large, i.e., of the order of $20\%$. The lifetimes hierarchy is
in
a good qualitative and even quantitative agreement with experiment except
for the $\Xi_c^{+}$ lifetime, which is somewhat smaller than the experimental
value.
Future measurements, especially measurements of the semileptonic branching
ratios for
$\Omega_c^{0}$, $\Xi_c^{+}$ and $\Xi_c^{0}$ should be decisive for the check of
this
approach.
{
}
\vskip2in
\end{quotation}
\vskip2in
\newpage

%
\setcounter{page}{1}
\noindent
{{\bf 1. Introduction}}
\vskip1cm

Weak decays of charmed and bottom hadrons [1-4] are particularly simple 
in the limit of infinite heavy quark mass. 

In reality, hadrons are bound states of heavy quark with light 
constituents (light quarks, gluons). The inclusion of soft 
degrees of freedom generates nonperturbative power corrections, 
an example of which is the destructive Pauli 
interference between the spectator light-quark and the light quark in 
$D^{+}$ meson 
coming from the decay of the heavy quark [5-7]. 

Inclusive hadronic decay rates and lifetimes are calculated some 
time ago [6-9]. It has been found that the overall picture for 
charmed hadron 
lifetimes is qualitatively satisfactory. Lifetime hierarchy has been 
predicted for charmed baryons [8,9], in qualitative agreement with 
present experiments [4]. 
It has also been shown [6,7] that the Pauli interference essentially 
lengthens the lifetime of the $D^{+}$ meson, thus being the main 
source of the $D^{0}-D^{+}$ lifetime difference. The systematic OPE 
brings in extra operators 
of dimension $5$, namely kinetic and chromomagnetic operators [10]. 
Their influence on charmed hadron lifetimes is rather moderate, because 
the main contribution comes from the four-quark operators.
Using the current quark mass, $m_c \sim 1.4 \, GeV$, the results for 
the charmed family are qualitatively good [1], provided one 
assumes that terms of higher dimension in the OPE drive 
the asymptotic (static) value of the decay constant, $F_{meson}$ 
close to the smaller value of the 
physical decay constant $f_{meson}$ only in meson decays, whereas in 
baryon decays that is not the case. In beauty decays, 
although everything was expected to work much better because of the 
large 
$b$-quark mass, a number of problems has remained unsolved [2,3,11]. 
In the first place, it is the discrepancy between the measured average 
semileptonic beauty 
meson branching ratio, which is somewhat smaller than the 
theoretical prediction. Second, the observed difference of the 
lifetimes of the $\Lambda_b$ baryon and of the $B$ meson is larger than 
expected. In view of these descrepancies, a 
radical phenomenological ansatz has been involved [11] with interesting 
consequences for charmed hadron decays. This approach, however, abandons 
the heavy-quark expansion and local duality, and requires the introduction 
[3] of a dominant destructive interference in the $\Xi_c^{+}$ decay and 
destructive $W$-anihilation in $D_s$ decays - the requirements that could 
hardly be satisfied if the four-quark operators were a dominant source of 
preasymptotic effects.

Recently, Voloshin has shown [12] that preasymptotic effects are 
largely present in semileptonic decays of charmed baryons. 
A significant enhancement [3,12] 
of $\Gamma_{SL}(\Xi_c)$ and $\Gamma_{SL}(\Omega_c)$ is expected 
relative to $\Gamma_{SL}(\Lambda_c)$ and $\Gamma_{SL}(D^{0})$ in 
order $1/m_c^3$ owing to the constructive Pauli interference in 
$\Gamma_{SL}(\Xi_c, \Omega_c)$ among the $s$-quarks.

In this paper we present theoretical predictions on the 
$\Lambda_c^{+}$, $\Xi_c^{+}$, $\Xi_c^{0}$ and $\Omega_c^{0}$ 
semileptonic 
branching ratios and lifetimes. The lifetimes have already been 
treated in [8,9]. Here, we extend previous calculations, in which 
preasymptotic effects in nonleptonic decay rates were basically 
attributed to the Cabbibo leading 
operators of dimension $6$. We make extension by including 
Voloshin's preasymptotic effects [12] 
in inclusive semileptonic decay rates. We calculate and add the Cabibbo 
subleading contributions since it was claimed in the literature [13] 
that they 
might be important in some cases owing to the statistical factors. 
We show that the constructive Pauli interference, which appears in 
the semileptonic $\Lambda_c^{+}$ decay at Cabbibo subleading level, is 
welcome because it enhances the theoretical value which is otherwise 
too small. Finally, we show that the inclusion of 
Voloshin's large preasymptotic effects in semileptonic decays does 
not destroy 
qualitative hierarchy of lifetimes, but improves it both 
qualitatively and quantitatively.
%

\noindent
\vskip2cm
\noindent
{{\bf 2. Preasymptotic effects in inclusive decays}}
\vskip1cm
  
The inclusive decay width of a hadron $H_c$, of the mass $M_{H_c}$,  
containing a $c$ quark can be 
written using the optical theorem as 
\be
\Gamma(H_c \rightarrow f)  = \frac{1}{2 M_{H_c}} 2\, Im \langle H_c|
\hat{T}|H_c\rangle\,,
\ee
i.e. as the forward matrix element of the imaginary part of the 
transition operator $\hat{T}$
\be
\hat{T} = i \int d^4x T\{ L_{eff}(x), L_{eff}^{\dagger}(0) \} .
\ee
The effective weak Lagrangian $L_{eff}(x)$ is given [1,3] as a sum of the 
semileptonic and nonleptonic part
\be
L_{eff} = L_{eff}^{SL} + L_{eff}^{NL}\,.
\ee

In the following we assume that the energy release in the decay of a $c$ quark 
is large enough so that momenta flowing through internal lines are also large 
and therefore justify the operator product expansion of the local operator (2). 
The result, widely discussed in the literature [1-3] is given by
\bea
\Gamma (H_c \rightarrow f) &=& \frac{G_{\rm F}^2 m_c^5}{192 \pi^3} |V|^2 
\frac{1}{2 M_{H_c}} \{ c_3^f \langle H_c|\overline{c} c|H_c\rangle + 
c_5^f \frac{ \langle H_c|\overline{c} g_s \sigma^{\mu \nu} G_{\mu \nu} c|
H_c\rangle }{m_c^2} \\ \nonumber
 &+& \sum_i c_6^f \frac{ \langle H_c| (\overline{c} 
\Gamma_i q)(\overline{q} \Gamma_i c)|H_c\rangle }{m_c^3} + O(1/m_c^4) + 
...\}\,. 
\eea
\\
Here $c_3^f$ and $c_5^f$ are coefficient functions which depend on the 
particular final state. \\
The coefficients $c_3^f$ are known at one-loop order and coefficients 
$c_5^f$ at tree level [10].


Let us calculate the semi-leptonic decay rates first. The main contribution is 
expected to come from the quark decay-type diagrams. When corrections
$O(m_c^{-2})$ are included, the contribution takes the form
\be
\Gamma_{SL}^{dec}(H_c) = \frac{G_F^2}{192\pi^3} m_c^5 
  (1 - \frac{1}{2}\frac{\mu_{\pi}^2(H_c)}{m_c^2}+
     \frac{1}{2}\frac{\mu_G^2(H_c)}{m_c^2}) F_1(x)\,.
\ee

Here $\mu_{\pi}^2(H_c)$ and $\mu_{\rm G}^2(H_c)$ parametrize the matrix 
elements of the kinetic energy and the chromo-magnetic operators, respectively. 
They can be determined from the spectrum of charmed heavy hadrons [14,15]. 
It turns out that only $\mu_G^2(\Omega_c^{0})$ is different from zero.

There is also the contribution of the dimension five operator
\be
\Gamma_{SL}^{G}(H_c) = \frac{G_F^2}{192\pi^3} m_c^5 (
-2 \frac{\mu_G^2(H_c)}{m_c^2}) F_2(x)\,.
\ee
 
We have included in the above expressions the phase space corrections $F_1(x)$ 
and $F_2(x)$ [16].

The semileptonic rate, as it has been shown by Voloshin [12], gets important 
and 
large contributions due to the preasymptotic effects. The 
result is given by
\be
\tilde{\Gamma}_{SL} = \frac{G_F^2}{12\pi} m_c^2 (4 \sqrt{\kappa}-1)|\psi(0)|^2\,.
\ee
Here, $|\psi(0)|$ is the baryon wave function at the origin and $\kappa$ is a 
correction due to the hybrid renormalization of the effective Lagrangian. 
 Hybrid renormalization is necessary since $|\psi(0)|^2$ is usually estimated 
in the effective quark models which are expected to make sense at the typical 
hadronic scales, $\mu = 0.5\sim 1\; GeV$. Therefore, it is necessary to 
evolve the 
effective Lagrangian from $m_c$ down to the scale $\mu$.

Total semileptonic rate is given by
\be
\Gamma_{SL}(H_c) = \Gamma_{SL}^{dec}(H_c) + \Gamma_{SL}^{G}(H_c) 
+ \Gamma_{SL}^{Voloshin}(H_c)\,,
\ee
where
\bea
\Gamma_{SL}^{Voloshin}(\Lambda_c^{+}) &=& s^2 \tilde{\Gamma}_{SL} \,,\nonumber\\
\Gamma_{SL}^{Voloshin}(\Xi_c^{+}) &=& \xi c^2 \tilde{\Gamma}_{SL} \,,\nonumber\\
\Gamma_{SL}^{Voloshin}(\Xi_c^{0}) &=& (\xi c^2 +s^2) \tilde{\Gamma}_{SL}\,, 
\nonumber\\
\Gamma_{SL}^{Voloshin}(\Omega_c^{0}) &=& \frac{10}{3} \xi c^2 
\tilde{\Gamma}_{SL}  \,.
\eea
Here $s^2$ and $c^2$ are abbreviations for $sin^2\theta_c$ and
$cos^2\theta_c$, and $\theta_c$ is the Cabibbo angle.

We have kept the Cabibbo-suppressed contributions because the 
preasymptotic effects are expected to be very large, and, therefore 
they might be a significant 
correction to $\Gamma_{SL}^{dec}(H_c)$. Also, we have introduced 
the parameter $\xi$ which is the ratio of the matrix elements of the operators
$(\bar{c}_L \gamma_{\mu} s_L)(\bar{s}_L \gamma^{\mu} c_L)$ and 
$(\bar{c}_L \gamma_{\mu} q_L)(\bar{q}_L \gamma^{\mu} c_L)$, where $q$ is $d$ 
or $u$ quark.
 $SU(3)$ symmetry-breaking effects, 
measured by $\xi$ are not expected to exceed $30 \%$. 
It is 
very difficult to reliably estimate the value of $\xi$, although certain 
hints can be made using different hadronic models. 
For example, the hadronic models used in [9] would suggest $\xi >1$, but 
in view of the fact that we lack reliable models such a conclusion might 
be premature. We shall not rely 
on such estimates in later discussions, but prefer to treat $\xi$ as a 
fitting parameter.

So far, only semileptonic branching ratio of $\Lambda_c^{+}$ has been measured
with reasonable accuracy
\be
BR(\Lambda^{+}_c \rightarrow e\,X) = (4.5 \pm 1.7 )\, \% \, .
\ee
By inspection of Eq.(9) one notes that except for $\Lambda_c^{+}$, 
all baryons receive potentially large Voloshin's contributions at the 
Cabibbo-leading level. Therefore, one expects, as pointed by Voloshin 
[12], significantly larger 
semileptonic ratio for $\Xi_c^{+}$, $\Xi_c^{0}$ and $\Omega_c^{0}$. 
Besides, the contribution to $\Lambda_c^{+}$ semileptonic branching 
ratio, although 
Cabbibo-suppressed by $sin^2\theta_c$, might be an important 
correction to the
decay diagram. This is welcome, since the decay diagram is not large 
enough to 
explain the experimental branching ratio [16].

The calculation of the nonleptonic decay rate closely follows the semileptonic 
ones. The lepton pair is substituted by a quark pair, and Wilson coefficients 
$c_{\pm}$ change their values because of the renormalization.
The contributions coming from the $c$-quark decay-type diagrams 
(including $O(m_c^{-2})$ corrections) and from the 
dimension five operator are of the form
\bea
\Gamma_{NL}^{dec}(H_c) &=& \frac{G_F^2}{192\pi^3}m_c^5
(c_{-}^2 + 2 c_{+}^2)(1 -
\frac{1}{2}\frac{\mu_{\pi}^2(H_c)}{m_c^2} +
\frac{1}{2}\frac{\mu_{G}^2(H_c)}{m_c^2}) F_1(x) \, ,\nonumber\\
\Gamma_{NL}^{G}(H_c) &=& -\frac{G_F^2}{192\pi^3}m_c^3(8 c_{+}^2 - 2 c_{-}^2)
\mu_{G}^2(H_c) \, .
\eea

The dominant contribution is expected to come from the preasymptotic effects. 
 They are given as
\bea
\Gamma^{ex} &=& \frac{G_F^2}{2\pi} m_c^2 [ c_{-}^2 + \frac{2}{3}(1 - \sqrt{\kappa})(c_{+}^2 - c_{-}^2)]\,|\psi(0)|^2\,, \nonumber \\
\Gamma^{int}_{-} &=& \frac{G_F^2}{2\pi} m_c^2 [ -\frac{1}{2} c_{+}(2 c_{-}-
c_{+}) - \frac{1}{6}(1 - \sqrt{\kappa})(5 c_{+}^2 + c_{-}^2-6 c_{+} c_{-})]
\,|\psi(0)|^2 \,,\nonumber \\
\Gamma^{int}_{+} &=& \frac{G_F^2}{2\pi} m_c^2 [ \frac{1}{2} c_{+}(2 c_{-}+
c_{+}) - \frac{1}{6}(1 - \sqrt{\kappa})(5 c_{+}^2 + c_{-}^2+6 c_{+} c_{-})]
\,|\psi(0)|^2 \, .
\eea

The result of the calculation of the nonleptonic rates is
\bea
\Gamma_{NL}(\Lambda_c^{+}) &=& \Gamma_{NL}^{dec}(\Lambda_c^{+}) + c^2 \Gamma^{ex} + \Gamma_{-}^{int} + s^2 \Gamma_{+}^{int} \,,\nonumber \\
\Gamma_{NL}(\Xi_c^{+}) &=& \Gamma_{NL}^{dec}(\Xi_c^{+}) + \xi s^2 \Gamma^{ex} 
+ \Gamma_{-}^{int} + \xi c^2 \Gamma_{+}^{int} \,,\nonumber \\
\Gamma_{NL}(\Xi_c^{0}) &=& \Gamma_{NL}^{dec}(\Xi_c^{-}) + (c^2+ \xi s^2) 
\Gamma^{ex} + (\xi c^2 + s^2) \Gamma_{+}^{int} \,,\nonumber \\
\Gamma_{NL}(\Omega_c^{0}) &=& \Gamma_{NL}^{dec}(\Omega_c^{0})  + 
\Gamma_{NL}^G(\Omega_c^{0}) 
+ \xi s^2 \frac{10}{3} \Gamma^{ex} + \xi c^2 \frac{10}{3} 
\Gamma^{int}_{+} \, .
\eea
Here we have not taken into account mass corrections, because they
are completely negligible.

By inspection of the results one sees that Cabibbo suppressed 
contribution 
only slightly changes the overall results. The right pattern depends 
on the 
value of $m_c$ and $|\psi(0)|^2$. However, for $\kappa =1$, 
$\Gamma_{+}^{int}$ is always larger than 
$\Gamma_{-}^{int}$. This conclusion holds even if hybrid logarithms are 
taken into account. Therefore, the $\Gamma_{+}^{int}$ will dominate the 
nonleptonic $\Xi_c^{+}$ decay rate, as far as $\xi\simeq 1$. Since 
experimentally $\Xi_c^{+}$ has the largest lifetime this rate should be 
relatively small.

For the determination of the baryon wave function we use estimates of the 
references [9,1], i.e the relation for the ratio of the squares of 
the meson and baryon wave functions
which is derived using the constituent quark model 
developed by De Rujula et al. [17]. There appear 'effective' quark 
masses, $m_c^{\ast} \simeq 1.5 GeV$ , $m_u^{\ast} \simeq 0.35 GeV$ 
and quarks are bound by a nonrelativistic potential which is modified 
by hyperfine interactions.

Following the approach of 
Ref.[1], in the expression for the baryonic wave function we shall use the 
static value $F_D$ instead of 
the physical decay constant $f_D$ for the reasons given below. This leads to
\be
|\psi^{\Lambda_c^{+}}(0)|^2 = \frac{3 (M_{\Sigma_c^{+}} - M_{\Lambda_c^{+}})}
{\mu_G^2(D)} m_u^{\ast} (\frac{1}{12} M_D F_D^2 \kappa^{-4/9})\, .
\ee
The importance of the value of 
$|\psi^{\Lambda_c^{+}}(0)|^2$ is obvious since the differences in 
decay widths/lifetimes are presumably generated mostly by the operators 
of dimension 6 (four-quark operators) and are therefore proportional to 
$F_D^2/m_c^2$, which indeed vanishes as $1/m_c^3$, since $F_D$ behaves 
as $m_c^{-1/2}$ for $m_c \rightarrow \infty$.

It has been argued [1], on a more intuitive 
basis, that in order to be consistent, one should use the static value 
$F_D$ in the calculations of the baryon decay. 
For meson decays, however, one should assume that the role of 
higher dimension terms is not negligible, and consequently, the 
physical (measured) constant $f_D$ should be used in calculations.

These arguments [1] are based on the fact that in meson 
decays one uses the factorization which necessarily brings into game 
the physical 
decay constant $f_D$, whereas in baryon decays this is not the case. Since 
there is no proof of this ansatz, one should take it as an attempt to 
disentangle the overall normalization of the mesonic matrix elements 
from the baryonic ones. 
In fact, it is known from previous calculations that, for example,
the destructive Pauli interference in the $D^{+}$ meson decay has 
reasonable 
values when the bag model wave functions are used [6], while the 
charmed baryon hierarchy was qualitatively well described by using 
nonrelativistic quark models [9]. In other words, in order 
to achieve agreement with experimental data, 
different normalization of matrix elements had to be used for 
meson and baryon decays.
\noindent
\vskip2cm
\noindent
{{\bf 3. Semileptonic branching ratios and lifetime hierarchy - 
results and discussions}}
\vskip1cm

Our choice of 'central' values of parameters roughly follows the set 
of values of Blok and Shifman [1].  
For $\Lambda_{QCD} =300\,MeV$, 
the Wilson coefficients are $c_{+} = 0.734$, $c_{-} = 1.856$. 

Our central value
for the charmed quark mass is $m_c = 1.4\,GeV$. However, in Table 1.
we show the results for $m_c = 1.35\,GeV$ for comparison. 
There is a controversy [15] over the value of $\mu_{\pi}^2$, 
which varies in the range [2]
\be
\mu_{\pi}^2 (B) \simeq -\lambda_1 = (0.3 \pm 0.2)\,GeV^2\, .
\ee
In our calculations we use the lower value, $\mu_{\pi}^2 = 0.1\,GeV^2$.
However, we also check that the larger value 
$\mu_{\pi}^2 = 0.5\,GeV^2$ only slightly
changes the result in semileptonic branching ratios, but has almost no 
effect on lifetimes. The chromomagnetic operator contributes only 
to $\Omega_c^{0}$ decays [1]; we use the value
\be
\mu_{G}^2 (\Omega_c^{0}) = 0.182 \, GeV^2 \,.
\ee

\begin{table}
\begin{center}
%
%
\begin{tabular}{|l|c|c|c|} \hline\hline
& \multicolumn{2}{c|}{RESULTS} &  \\
& \multicolumn{2}{c|}{
   ($\xi = 1,\,
   \Lambda_{QCD} = 300\, MeV$,\,$\mu = 1 \,GeV$)}
& EXP. DATA \\ \cline{2-3}
& $\;\;\;\;m_c = 1.4\, GeV \;\;\;\;$
& $\;\;\;\;m_c = 1.35\, GeV \;\;\;\;$ & [4]   \\ \hline\hline
&  \multicolumn{2}{c|}{\em Lifetimes in units $10^{-13}$ s} & \\ \hline\hline
$\tau (\Lambda^+_c)$ & 2.03 & 2.18 & $2.06 \pm 0.12$ \\ \hline
$\tau (\Xi^{+}_c)$ & 2.42 & 2.70  & $3.50 \pm 0.70$ \\ \hline
$\tau (\Xi^0_c)$ & 0.87 & 0.94 & $0.98 \pm 0.23$ \\ \hline
$\tau (\Omega^0_c)$ & 0.58 & 0.63 & $0.64 \pm 0.20$ \\ \hline\hline
&  \multicolumn{2}{c|}{\em Lifetime ratios} & \\ \hline\hline
$\tau (\Xi^{+}_c)/\tau (\Lambda^+_c)$ & 1.19 & 1.23 & $1.69 \pm 0.35$ \\ \hline
$\tau (\Xi^{0}_c)/\tau (\Lambda^+_c)$ & 0.43 & 0.43  & $0.47 \pm 0.11$ \\ \hline
$\tau (\Omega^0_c)/\tau (\Lambda^+_c)$ & 0.28 & 0.29 & $0.31 \pm 0.09$ \\ \hline
$\tau (\Xi^{+}_c)/\tau (\Xi^0_c)$ & 2.76 & 2.87 & $3.57 \pm 1.10$ \\ \hline
 &  \multicolumn{2}{c|}{\em Semileptonic decay rates in units $ps^{-1}$} &
 \\ \hline\hline
$\Gamma_{SL}(\Lambda^+_c)$ & 0.209 & 0.177 & $0.225 \pm 0.085$ \\ \hline
$\Gamma_{SL}(\Xi^{+}_c)$ & 1.010 & 0.918 & \rule[0.01in]{0.3in}{0.01in}
\\ \hline
$\Gamma_{SL}(\Xi^0_c)$ & 1.053 & 0.958 & \rule[0.01in]{0.3in}{0.01in}
       \\ \hline
$\Gamma_{SL}(\Omega^0_c)$ & 2.954 & 2.718 & \rule[0.01in]{0.3in}{0.01in}
      \\ \hline\hline
& \multicolumn{2}{c|}{\em Semileptonic branching ratios in $\%$} &
       \\ \hline\hline
$BR_{SL}(\Lambda^+_c)$ & 4.2 & 3.8 & $4.5 \pm 1.7$
  \\ \hline
$BR_{SL}(\Xi^{+}_c)$ & 24.5 & 24.8 & \rule[0.01in]{0.3in}{0.01in}
  \\ \hline
$BR_{SL}(\Xi^0_c)$ & 9.2 & 9.0 & \rule[0.01in]{0.3in}{0.01in}
  \\ \hline
$BR_{SL}(\Omega^0_c)$ & 17.4 & 17.3 & \rule[0.01in]{0.3in}{0.01in}
  \\ \hline\hline
\end{tabular}
\caption{Predictions for semileptonic branching ratios and lifetimes of
charmed baryons given for two values of the charmed quark mass $m_c$.}
\end{center}
\end{table}

Following [1], we use the following input to obtain the central value 
in Eq.(14): $F_D = 400 \, MeV$, $m_u^{\ast} = 350\, MeV$, $M_{\Sigma_c
^{\ast}} - M_{\Lambda_c^{+}} = 400\, MeV$ (static value) and 
$M_D = 1870\, MeV$. 
For $\mu_{G}^2(D)$, we use the value $\mu_{G}^2(D) = 0.4\, GeV^2$. 
Then, for $\kappa =1$, Eq.(14) gives our central value for the baryon 
wave function
\be
|\psi^{\Lambda_c^{+}}(0)|^2_{\mu =m_c} = 0.0262\, GeV^3 \,.
\ee

Our numerical results are presented in Tables 1-2. The left set of 
numbers in the Tables are numerical results obtained for our central 
values discussed before, and for $m_c = 1.4\,GeV$, $m_s = 150\,MeV$, 
$\mu = 1\,GeV$ and 
$\xi = 1$. The agreement with available experimental data is very good, 
except for the lifetime of $\Xi_c^{+}$, where theoretically predicted 
value is smaller than in experiment (see Table 1). The same problem 
with $\Xi_c^{+}$ persists if one calculates the ratio of lifetimes 
(in such a way significantly reducing the uncertainty coming from the 
wave-function value). However, the semileptonic branching ratio for 
$\Lambda_c^{+}$ is in excellent agreement with the experimental value,
showing clearly that preasymptotic effects of Voloshin's type, 
although at the Cabibbo suppressed level, significantly improve the 
theoretical value.

For $\Lambda_{QCD}= 200\,MeV$ the results are almost 
not affected except the $\Lambda_c^{+}$ lifetime which grows by $20\%$, 
becoming so unpleasantly large. However, one should keep in mind that 
such modest discrepances are expected since in charmed baryon decays 
we are far away from the asymptotic limit.

\begin{table}
\begin{center}
%
\begin{tabular}{|l|c|c|c|} \hline\hline
& \multicolumn{2}{c|}{RESULTS} &  \\
& \multicolumn{2}{c|}{
 ($m_c = 1.4\, GeV,
 \Lambda_{QCD} = 300\, MeV$,\,$\mu = 1 \,GeV$)}
 & EXP. DATA \\ \cline{2-3}
& $\;\;\;\;\;\;\;\;\;\;\;\;\;\;\;\;\;\;\;\xi =1 \;\;\;\;\;\;\;\;\;\;\;\;\;$
& $\xi =0.75$ & [4]   \\ \hline\hline
&  \multicolumn{2}{c|}{\em Lifetimes in units $10^{-13}$ s} & \\ \hline\hline
$\tau (\Lambda^+_c)$ & 2.03 & 2.03 & $2.06 \pm 0.12$ \\ \hline
$\tau (\Xi^{+}_c)$ & 2.42 & 3.41  & $3.50 \pm 0.70$ \\ \hline
$\tau (\Xi^0_c)$ & 0.87 & 0.98 & $0.98 \pm 0.23$ \\ \hline
$\tau (\Omega^0_c)$ & 0.58 & 0.76 & $0.64 \pm 0.20$ \\ \hline\hline
e
e
& \multicolumn{2}{c|}{\em Semileptonic branching ratios in $\%$} &
\\ \hline\hline
$BR_{SL}(\Lambda^+_c)$ & 4.2 & 4.2 & $4.5 \pm 1.7$
 \\ \hline
$BR_{SL}(\Xi^{+}_c)$ & 24.5 & 27.2 & \rule[0.01in]{0.3in}{0.01in}
   \\ \hline
$BR_{SL}(\Xi^0_c)$ & 9.2 & 8.2 & \rule[0.01in]{0.3in}{0.01in}
   \\ \hline
$BR_{SL}(\Omega^0_c)$ & 17.4 & 17.3 & \rule[0.01in]{0.3in}{0.01in}
    \\ \hline\hline
\end{tabular}
\caption{Predictions for charmed baryons in dependence of
the parameter $\xi$.}
\end{center}
\end{table}

In Table 1 we also display results of the calculations 
for the smaller value of the current quark mass, $m_c = 1.35\,GeV$. 
Again, the results are not very sensitive to this variation, although 
the agreement with experiment is slightly improved, especially for 
$\Xi_c^{+}$ decays. 

In Table 2 we have presented the results of calculations for the 
specific choice of the parameter $\xi$, $\xi = 0.75$, compared with 
the results obtained for $\xi =1$. As discussed above, we allow 
$\xi$ to have a value different from $1$, treating it as a 
free parameter. Fitting $\xi$ roughly to the value needed to bring 
the $\Xi_c^{+}$ lifetime into agreement with experiment gives 
$\xi \simeq 0.75$. Fitting $\xi$ basically means to fit the 
constructive interference term $\Gamma_{+}^{int}$ for Cabibbo favoured  
decays. The simple fit of one lifetime would of course, not make much 
progress. However, $\Gamma_{+}^{int}$ enters also the decay rates of 
other baryons, $\Xi_{c}^{0}$ and $\Omega_c^{0}$, at the same Cabibbo 
level. Besides, the same factor $\xi$ enters Voloshin's contributions 
to semileptonic decay rates, again at the same Cabibbo level. Therefore,
any trivial fit to $\Xi_c^{+}$ could at the same time worsen the results 
for other particles. However, interestingly enough, here this is not the 
case. Although the fit $\xi = 0.75$ brings the $\Xi_c^{+}$ lifetime 
into perfect agreement between theory and experiment, it does not spoil
the agreement between theory and experiment for both semileptonic BR's
and lifetimes of other particles.

Next we study the dependence of lifetimes on the square of the baryon
 wave function $|\psi(0)|^2$ and show the results in Fig.1 and 2. We 
vary the value $|\psi(0)|^2$ inside a factor of 2 in the range 
\be
0.018\, GeV^3 \leq |\psi^{\Lambda_c^{+}}(0)|^2_{\mu = m_c} 
\leq 0.034 \,GeV^3\,,
\ee
with a central value given by (17). 

\figgeps{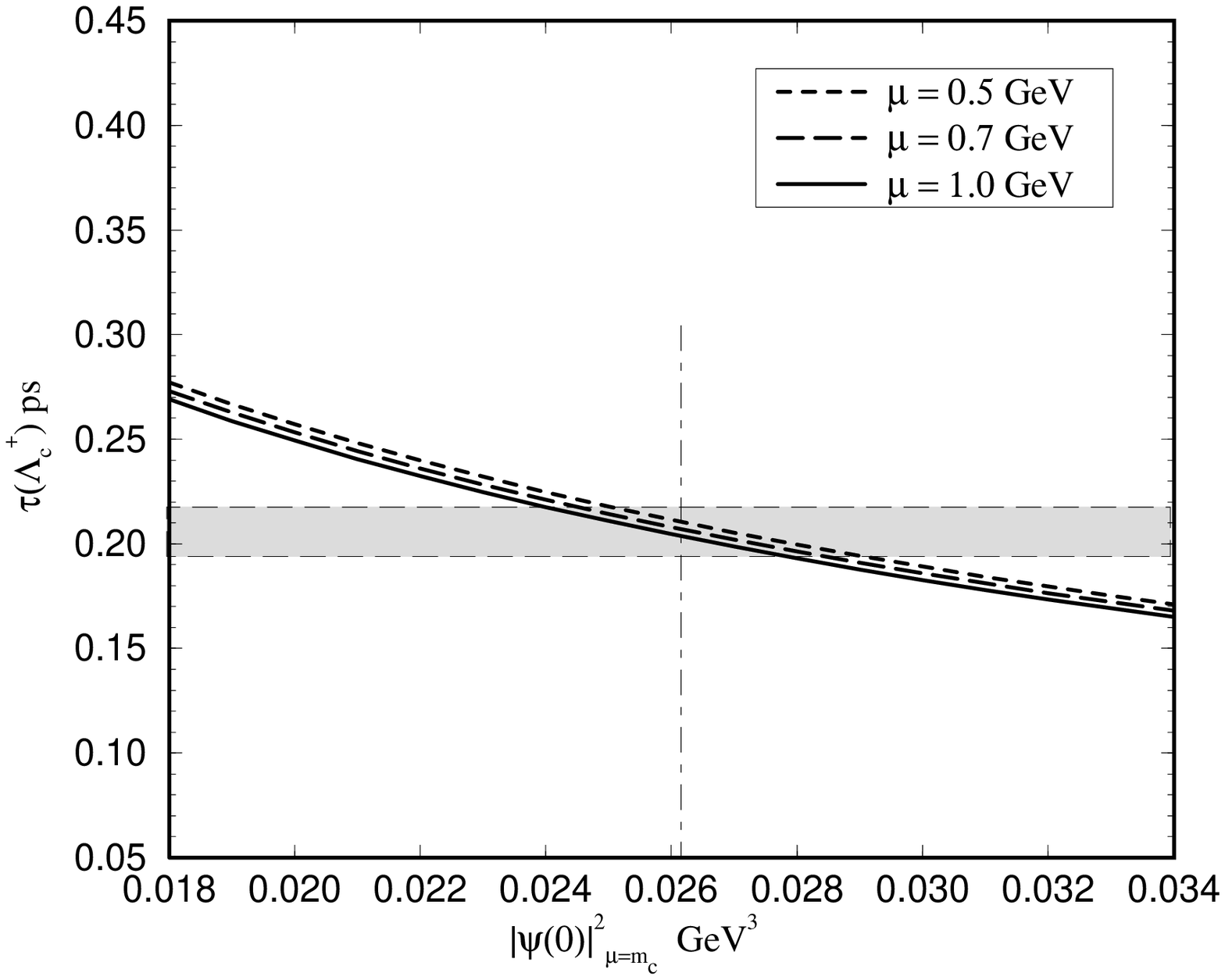}{8}{8}{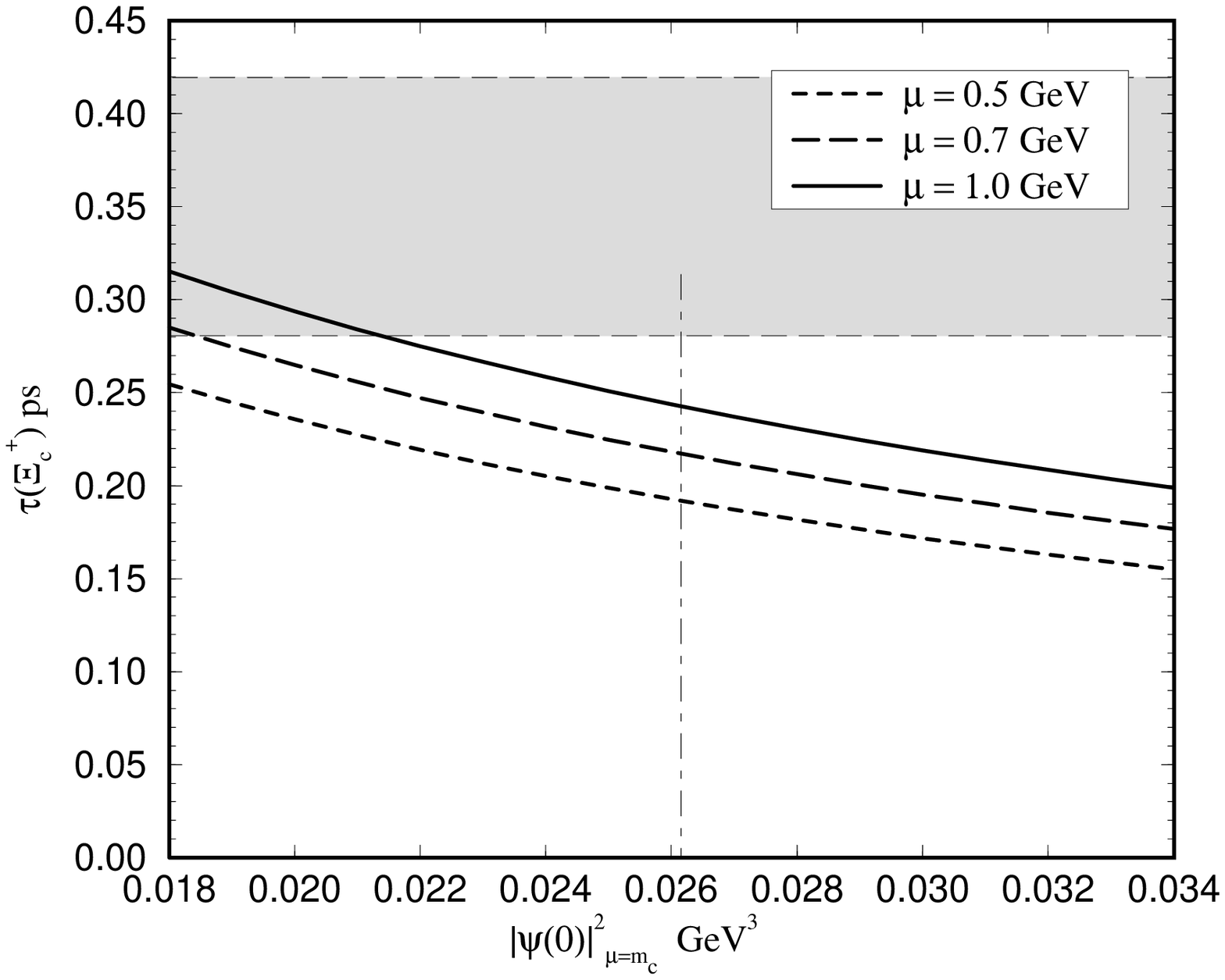}{8}{8}{ Lifetimes of $\Lambda_c^{+}$ and
$\Xi_c^{+}$ as a function of the square of the
baryon wave function $|\psi(0)|^2$,
given for three values of the hybrid renormalization point $\mu$.
The shaded areas are the experimentally allowed regions.
The dot-dashed vertical line is a value of
$|\psi(0)|^2$ used in Tables 1 to 2. The results are obtained using
$m_c = 1.4 \,GeV$, $\Lambda_{QCD} = 300\, MeV$, $\mu_{\pi}^2
= 0.1\, GeV^2$.
}{}

It is interesting to note that the lifetimes of $\Lambda_c^{+}$ and 
$\Xi_c^{+}$ are 
very sensitive to the value of $|\psi^{\Lambda_c^{+}}(0)|^2$, Fig.1.
 On the other hand, the other decays are not so 
sensitive, and are consistent with experiment even for the lower
value of $|\psi^{\Lambda_c^{+}}(0)|^2$ given in (18), Fig.2. Therefore, one 
could easily bring the lifetimes of $\Xi_c^{+}$, $\Xi_c^{0}$ and 
$\Omega_c^{0}$ to agreement with experiment simply by using a smaller 
value of $|\psi^{\Lambda_c^{+}}(0)|^2$. This would enlarge the 
$\Lambda_c^{+}$ lifetime and introduce a descrepancy between theory
and experiments. However, our analysis of lifetimes, as discussed
above, shows that the $\Xi_c^{+}$ lifetime exhibits a peculiar behavior
- strong $\mu$-dependence and strong $|\psi(0)|^2$-dependence. Furthermore, 
the central value of $|\psi^{\Lambda_c^{+}}(0)|^2$, given in (36), 
gives a good semileptonic branching ratio of $\Lambda_c^{+}$, which 
we discuss next.

\figgeps{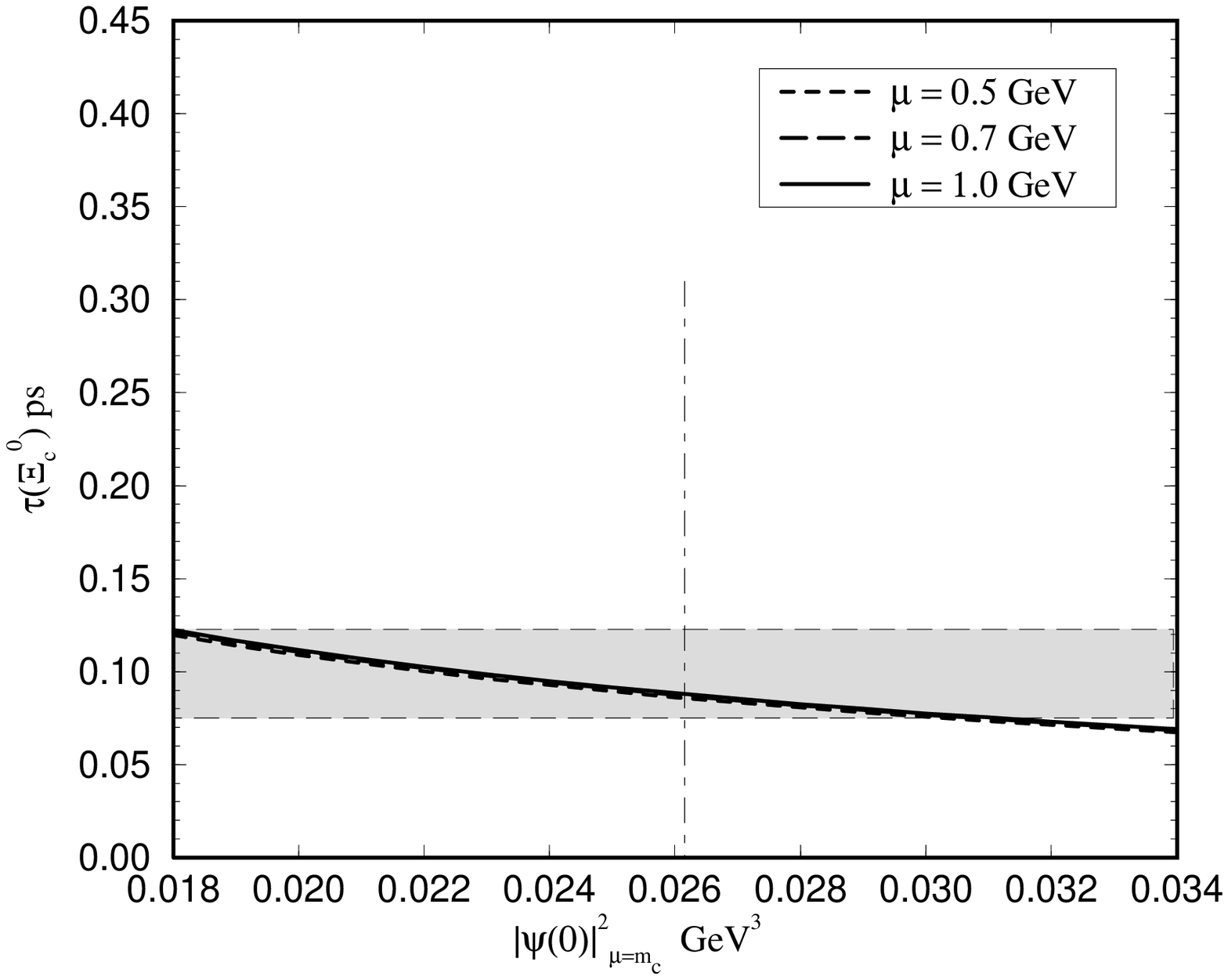}{8}{8}{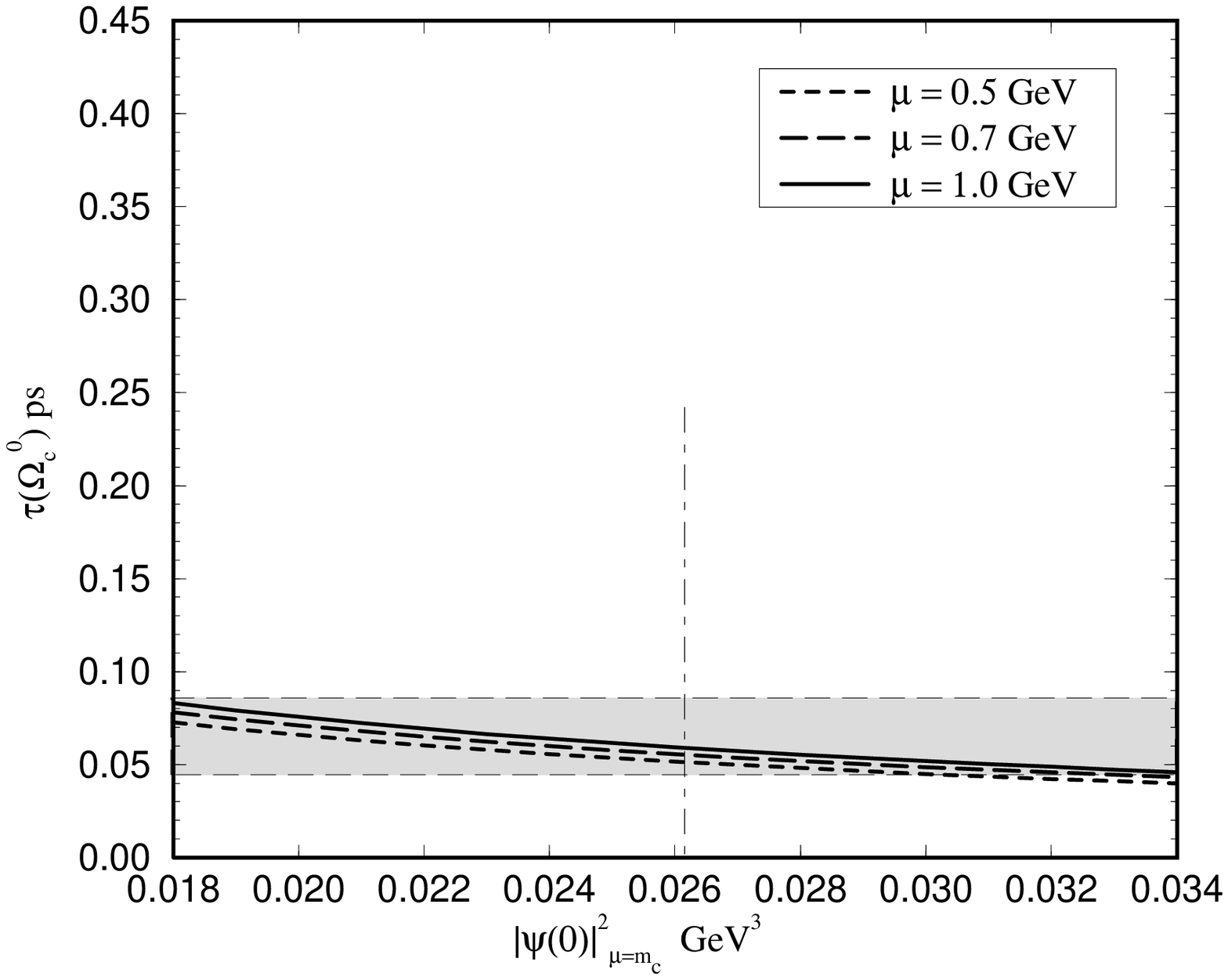}{8}{8}
{Lifetimes of $\Xi_c^{0}$ and $\Omega_c^{0}$
 as a function of the square of the baryon wave
 function $|\psi(0)|^2$ given for
 three values of the hybrid renormalization point $\mu$.
 The shaded areas are the experimentally allowed regions.
 The dash-dotted vertical line is a value of
 $|\psi(0)|^2$ used in Tables 1 to 2. The results are obtained using
 $m_c = 1.4 \,GeV$, $\Lambda_{QCD} = 300\, MeV$, $\mu_{\pi}^2 =
 0.1\, GeV^2$, $\mu_G^2(\Omega_c^{0}) = 0.182\, GeV^2$.}{}

As discussed in the preceding section, baryons receive Voloshin's large interference contributions at the Cabibbo leading level. Their role is 
obvious from Table 1, where a certain hierarchy of semileptonic BR's 
is strongly pronounced:
\be
BR_{SL}(\Lambda_c^{+}) < BR_{SL}(\Xi_c^{0}) < BR_{SL}(\Omega_c^{0}) 
< BR_{SL}(\Xi_c^{+})\,.
\ee
It is in the numerical range from 4.5 to 25 percent. We consider 
prediction (19) as a crucial test of the approach presented here, for 
the following reasons: 
Voloshin's interference effects, being proportional to $|\psi(0)|^2$
are necessarily large, because one needs a large $|\psi(0)|^2$ in 
order to reproduce experimental values of lifetimes. If one finds 
experimentally that all semileptonic BR's are of the order of $BR_{SL}
(\Lambda_c^{+})$, this will mean that the interference effect in 
semileptonic decays is negligible, and that, therefore, the 
preasymptotic effects in (12) are unlikely to be responsible for the 
experimentally evident hierarchy of lifetimes.

In addition to this very clear prediction, Voloshin's interference 
effect helps to improve the theoretical value of the $\Lambda_c^{+}$ 
semileptonic branching ratio. It appears at the Cabibbo suppressed 
level and acts as a correction to the main contribution coming 
from the decay diagram. It has been known for a long time 
that the quark decay mechanism cannot explain the 
semileptonic branching ratio of $\Lambda_c^{+}$, if one uses the 
current quark mass $m_c \sim 1.4\,GeV$ in (5). An effective mass 
of the order $1.6-1.7 \,GeV$ is actually needed [1-3,11]. 

In Fig.3 we show the semileptonic branching ratio for $\Lambda_c^{+}$
 as a function of $|\psi(0)|^2$, for two cases, namely with and 
without finite $\alpha_s$-corrections. 
It is quite clear that the interference effect is welcome, because it 
brings the theoretical value close to experiment. 
In all calculations, presented in Tables 1-2, 
the finite $\alpha_s$-corrections have not been taken into account 
because they are not known for all kinds of preasymptotic contribution. 
However, the right figure in Fig.3 shows the semileptonic branching 
ratio with finite $\alpha_s$-corrections included. Obviously, without 
the interference effect the theoretical value is outside the 
experimentally allowed region. 

\figgeps{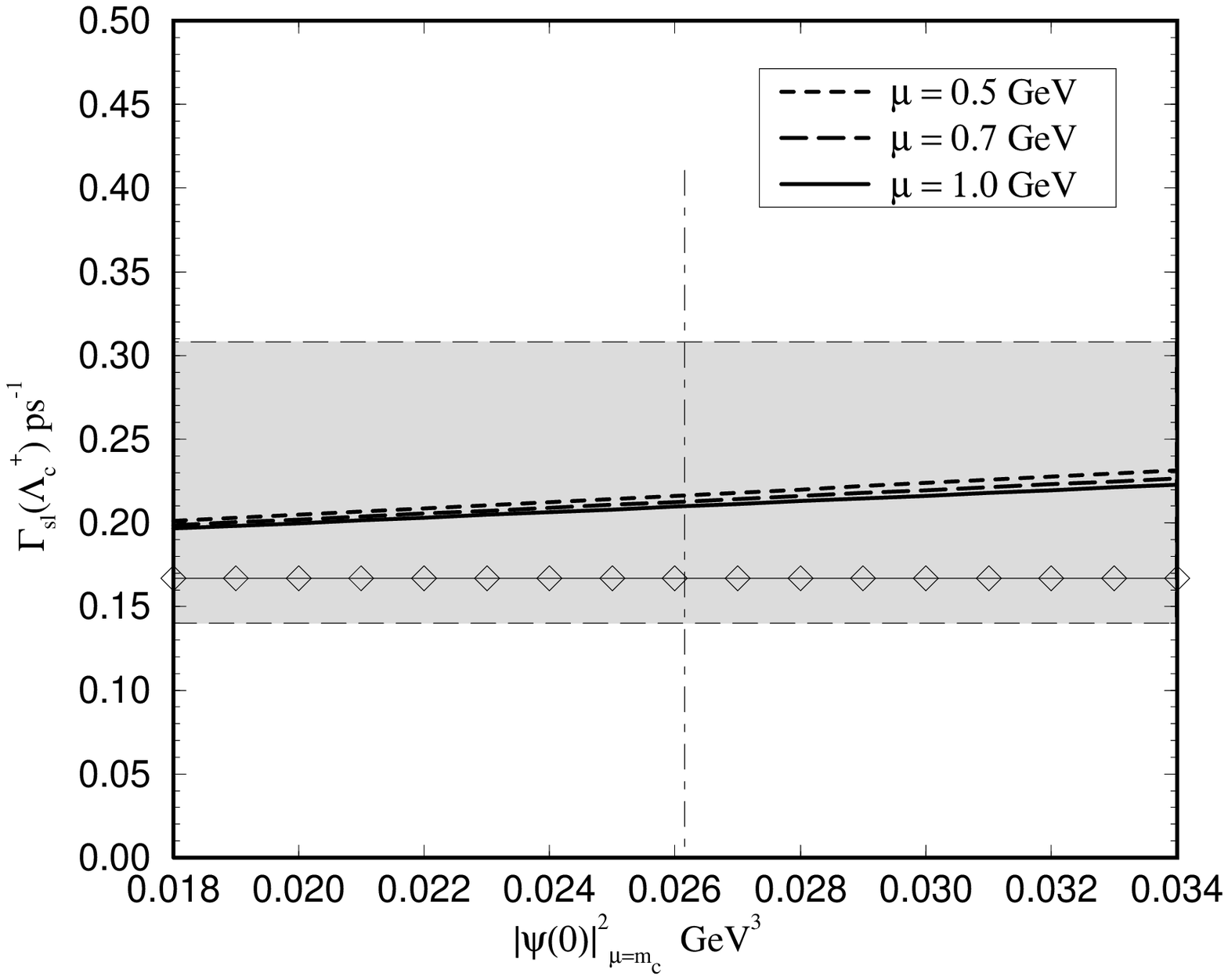}{8}{8}{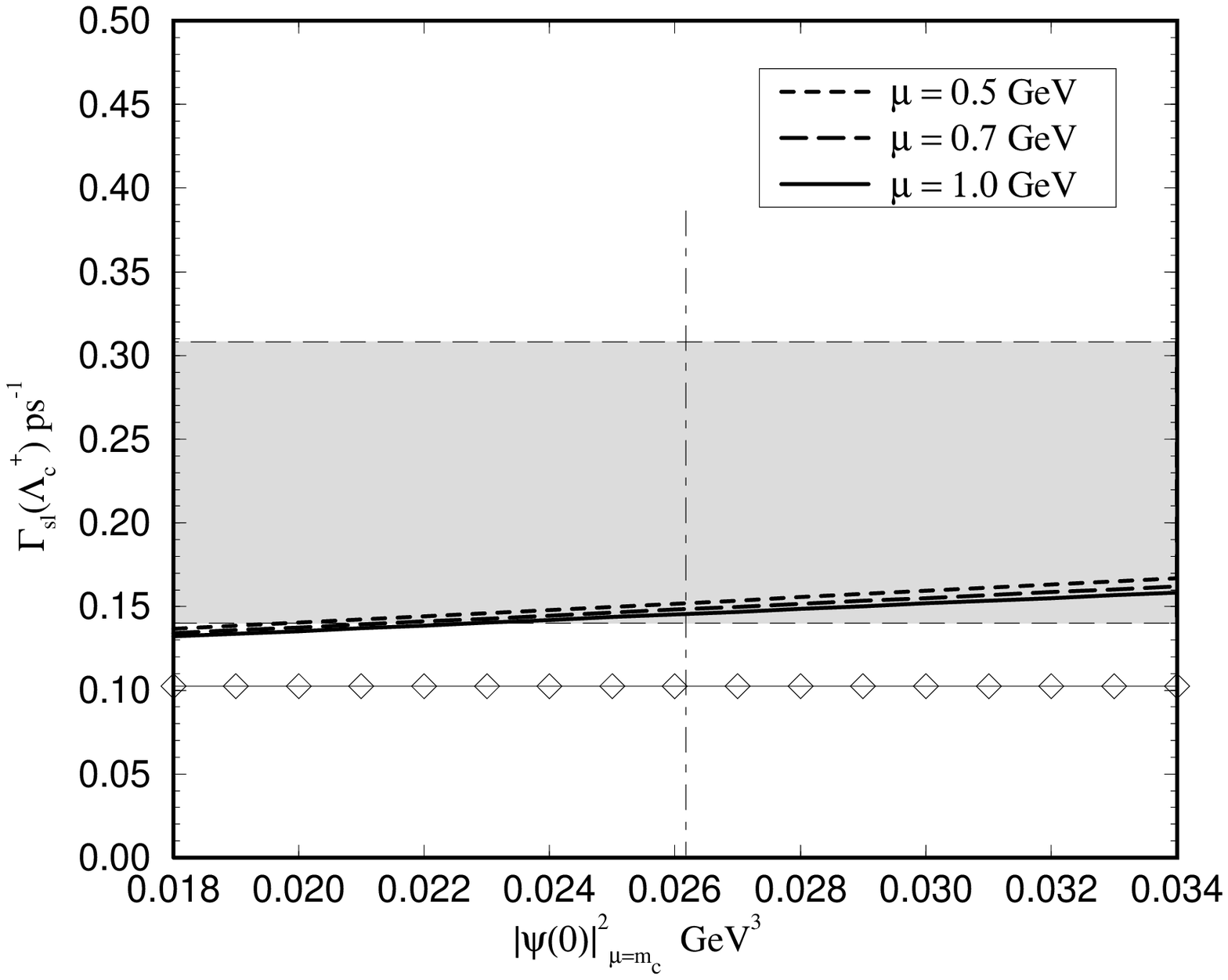}{8}{8}{Semileptonic $\Lambda_c^{+}$-decay rate
for three values
of the hybrid renormalization point $\mu$ is given as a function
of the square of the baryon
wave function $|\psi(0)|^2$. On the left picture finite
$\alpha_s$-contributions are not included.
The vertical dot-dashed line denotes the 'central'
value in (17). The shaded area is the experimentally
allowed region. $\diamondsuit$ denotes the semileptonic decay rate without
Voloshin's contributions. The
other parameter values used are $m_c = 1.4
\,GeV$, $\Lambda_{QCD} = 300\, MeV$, $\mu_{\pi}^2 = 0.1\, GeV^2$.
}{}
 
\noindent
\vskip2cm
\noindent
{{\bf 4. Conclusions}} 
\vskip1cm

In this paper we have performed an analysis of inclusive semileptonic branching
ratios and lifetimes for the charmed baryon family. New ingredients in this 
analysis are the inclusion of preasymptotic effects in semileptonic decays 
and the inclusion of Cabibbo suppressed contributions. In the calculations 
we have used the input parameters determined by QCD, 
thus following
the approach of Blok and Shifman [1], i.e. we have avoided the introduction 
of "effective" parameters, such as effective charmed quark mass, effective 
hadron mass instead of quark mass, etc. In this way we have tried to test 
quark-hadron duality using our present knowledge of OPE and QCD 
up 
to the level of introducing the baryon wave function, $|\psi^{\Lambda_c^{+}}
(0)|$ as a measure of the strength of the dominant preasymptotic effects. 
Having fixed the charmed quark mass to be approximately $m_c \sim 1.4\,GeV$, 
the semileptonic branching ratios and lifetimes of charmed baryons depend 
essentially on the square of the baryon wave function, $|\psi^{\Lambda_c^{+}}
(0)|^2$, which, in spirit of the above considerations may be regarded as a 
fitting parameter. It is a pleasent discovery that a rough fit of 
$|\psi^{\Lambda_c^{+}}(0)|^2$ agrees very well with the Blok-Shifman estimate 
[1].

Our analysis leads to the following conclusions:
\\
i) The inclusion of Voloshin's large preasymptotic effects leads to following 
predictions: The semileptonic branching ratios of $\Xi_c^{+}$, $\Xi_c^{0}$, 
and $\Omega_c^{0}$ are significantly larger than the semileptonic branching 
ratios of $\Lambda_c^{+}$ with a hierarchy already given in Eq.(19):
\[BR_{SL}(\Lambda_c^{+}) < BR_{SL}(\Xi_c^{0}) < BR_{SL}(\Omega_c^{0}) < BR_{SL}(\Xi_c^{+}) \,. \]
The inclusion of the Cabibbo suppresed interference effect in the semileptonic 
decay rate of $\Lambda_c^{+}$ enhances it and brings the branching ratio to 
agreement with experiment.
\\
ii) The change in semileptonic decay rates which is due to interference effects
significantly helps to obtain very good qualitative and even quantitive 
results for the lifetimes with the same hierarchy 
\be
\tau(\Omega_c^{0}) < \tau(\Xi_c^{0}) < \tau(\Lambda_c^{+}) < \tau(\Xi_c^{+})\,,
\ee
as predicted in [8,9]. The predicted lifetime of $\Xi_c^{+}$ appears to be 
somewhat smaller than the experimental value. We do not consider that as a 
problem, since we are far away from the asymptotic limit and it would be 
premature to expect that higher order terms in OPE are really negligible.
\\
iii) Concerning the results as obtained and shown in Tables 1-2, one may 
conclude that quark-hadron duality works suprisingly well for the 
charmed baryon family.

After completion of this work we have learned of a recent paper of Cheng [18]
where Voloshin's type of corrections were considered in a different context, 
in a more phenomenological way, introducing the effective charmed-quark mass,
substitution of the universal quark mass by the particular physical 
hadron mass, etc. 
We believe that our results, compared with the results of Cheng [18], are 
more consistent and more reliable.
\vskip0.81in
\noindent
{\bf Acknowledgement}
\vskip0.3in
This work was supported by the Ministry of Science and Technology of the 
Republic of Croatia under the contract Nr. 00980102.

\newpage
\noindent
{\bf References}
\noindent
\vskip1cm
\noindent
\begin{enumerate}
\item B. Blok and M. Shifman: "Lifetimes of Charmed Hadrons Revised - Facts and Fancy." in Proc. of the Workshop on the Tau-Charm Factory, p.247, 
Marbella, Spain, 1993, eds. J. Kirkby and R. Kirkby (Editions Frontiers, Gif-
sur-Yvette, 1994), TPI-MINN-93/55-T, UMN-TH-1227/93.
\\
\item M. Neubert: "$B$ Decays and the Heavy-Quark Expansion", preprint 
CERN-TH/97-24 and hep-ph/9702375, to appear in 'Heavy Flavors II', 
eds. A.J. Buras and M. Lindner.
\\
\item I.I. Bigi: "Heavy Quark Expansions for Inclusive Heavy-Flavor Decays and the Lifetimes of Charm and Beauty Hadrons", invited lecture given at HQ96 'Heavy 
Quarks at Fixed Target', St. Goar, Germany, Oct. 3.-6., 1996, UND-HEP-
96-BIG06 and hep-ph/9612293.
\\
\item Review of Particle Properties: Phys. Rev. D54 (1996).
\\
\item B. Guberina, S. Nussinov, R.D. Peccei and R. R\"{u}ckl: Phys. Lett. B89 (1979) 111.
\\
\item N. Bili\'c, B. Guberina and J. Trampeti\'c: Nucl. Phys. B248 (1984) 261.
\\
\item M. Shifman and M. Voloshin: Yad. Fiz. 41 (1985) 463 [ Sov. J. Nucl. Phys. 
 41 (1985) 120].
\\
\item M. Shifman and M. Voloshin: ZhETF 91 (1986) 1180 [ JETP  64 (1986) 
698].
\\
\item B. Guberina, R. R\"{u}ckl and J. Trampeti\'c: Z. Phys. C33 (1986) 297.
\\
\item M. Neubert: Phys. Rep. 245 (1994) 259; Int. J. Mod. Phys.  A11 (1996)4173.
\\
\item G. Altarelli, G. Martinelli, S. Petrarca and F. Rapuano: Phys. Lett. B382 (1996) 409.
\\
\item M. Voloshin: Phys. Lett. B385 (1996) 369.
\\
\item V. Gupta and K.V.L. Sarma: Int. J. Mod. Phys. A5 (1990) 879.
\\
\item I. Bigi, M. Shifman and N. Uraltsev: "Aspects of Heavy Quark Theory", 
TPI-MINN-97/02-T, UMN-TH-1528-97, UND-HEP-97-BIG01 and hep-ph/9703290.
\\
\item M. Neubert: "Theory of Inclusive $B$ Decays", invited talk at the $4^{th}$ KEK Topical Conference on Flavor Physics, KEK, Japan, Oct. 29-31, 1996, 
CERN-TH/9-19 and hep-ph/9702310.
\\
\item I. Bigi: "The QCD Perspective on Lifetimes of Heavy-Flavour Hadrons", preliminary version of the Physics Report paper, UND-HEP-95-BIG02 and 
hep-ph/9508408.
\\
\item A. de Rujula, H. Georgi and S. Glashow: Phys. Rev. D12 (1975) 147; 
J.L. Cortes and J. Sanchez-Guillen: Phys. Rev. D24 (1981) 2982.
\\
\item H.Y. Cheng: "A Phenomenological Analysis of Heavy Hadron Lifetimes", 
IP-ASTP-02-97 and hep-ph/9704260.
\end{enumerate}

\end{document}